\documentclass{ifacconf}
\usepackage{graphicx}      % include this line if your document contains figures
\usepackage{natbib}        % required for bibliography

\usepackage{dsfont} 
\usepackage{amsmath} 
\usepackage{amssymb}
\usepackage{color}
\usepackage{url}
\usepackage{afterpage}
\usepackage{comment}
\usepackage{mdframed}
\usepackage{tcolorbox}
\usepackage{changepage}
\usepackage{mathrsfs}
\usepackage{comment}
\usepackage{caption}
\usepackage{subcaption}
\usepackage{cuted}
\usepackage{etoolbox}
\usepackage{mathtools}
\usepackage[mathscr]{euscript}
\usepackage{subcaption}
\usepackage{caption}
\usepackage{algorithm,algcompatible}
\usepackage{enumerate}

\newtheorem{remark}{Remark}

\DeclareMathOperator*{\argmax}{argmax}

\begin{document}
\begin{frontmatter}

\title{A Data-Driven Automatic Tuning Method for MPC under Uncertainty using Constrained Bayesian Optimization} 
% Data-Driven Automatic Tuning of MPC under Uncertainty using Constrained Bayesian Optimization
% A Data-Driven Auto-Tuning Approach for MPC under Uncertainty using Constrained Bayesian Optimization
% Goal-Oriented Model Learning using MPC-Embedded Constrained Bayesian Optimization
% Title, preferably not more than 10 words.

%\thanks[footnoteinfo]{Sponsor and financial support acknowledgment
%goes here. Paper titles should be written in uppercase and lowercase
%letters, not all uppercase.}

\author[First,Second]{Farshud Sorourifar} 
\author[Second]{Georgios Makrygirgos} 
\author[Second]{Ali Mesbah}
\author[First]{Joel A. Paulson}

\address[First]{Department of Chemical and Biomolecular Engineering, The Ohio State University, Columbus, OH 43210, USA}   
\address[Second]{Department of Chemical and Biomolecular Engineering, University of California, Berkeley, CA 94720, USA}

\begin{abstract}                % Abstract of not more than 250 words.
The closed-loop performance of model predictive controllers (MPCs) is sensitive to the choice of prediction models, controller formulation, and tuning parameters. However, prediction models are typically optimized for prediction accuracy instead of performance, and MPC tuning is typically done manually to satisfy (probabilistic) constraints. In this work, we demonstrate a general approach for automating the tuning of MPC under uncertainty. In particular, we formulate the automated tuning problem as a constrained black-box optimization problem that can be tackled with derivative-free optimization. We rely on a constrained variant of Bayesian optimization (BO) to solve the MPC tuning problem that can directly handle noisy and expensive-to-evaluate functions. The benefits of the proposed automated tuning approach are demonstrated on a benchmark continuously stirred tank reactor example.
\end{abstract}


\end{frontmatter}
%===============================================================================

%%%%%%%%%%%%%%%%%%%%%%%%%%%%%%
\section{Introduction}
\label{sec:intro}

Model predictive control (MPC) is one of the most widely used methods for the control of constrained multivariable systems \citep{rawlings2009model}. The closed-loop performance of MPC strongly depends on (i) the quality of its underlying process model used to make dynamic predictions; (ii) the formulation of the objective and constraints; and (iii) the choice of several tuning parameters (e.g., prediction horizon, weights in cost function, and constraint backoff terms) \citep{garriga2010model,paulson2018nonlinear}. A major challenge in MPC tuning arises from the non-trivial relationships between the tuning parameters and the closed-loop control performance and constraint satisfaction \citep{lu2020mpc}. As such, MPC tuning via trial-and-error or other heuristic strategies may require a significant number of closed-loop simulations, which can quickly become prohibitive especially when system uncertainties are considered.

Recently, there has been a renewed interest in automated strategies for controller tuning using Bayesian optimization (BO); e.g., see \citet{berkenkamp2016safe,bansal2017goal,neumann2019data,forgione2019efficient,khosravi2020performance,lu2020mpc}. BO has emerged as a powerful derivative-free method for optimizing black-box functions in various applications \citep{shahriari2015taking}, most notably for hyperparameter selection of machine learning algorithms \citep{snoek2012practical}. BO is deemed particularly useful for solving MPC tuning problems since it can accommodate a mixture of continuous and discrete decision variables and also overcome the limitations of alternative derivative-free optimization methods such as genetic algorithms and particle swarm optimization \citep{garriga2010model}.    

In any model-based control approach, the chosen model plays a pivotal role in the design of the controller; however, identification of the model has been traditionally separated from controller design (even in the context of automated tuning strategies). An alternative idea that has been gaining popularity in recent years is to treat the identification process as a ``tuning parameter'' that should take into account the intended control application. The identification for control (I4C) rationale has been heavily studied in the context of fixed-order controllers for linear time-invariant systems \citep{gevers2005identification}. Recently, the I4C methodology was extended to MPC in \citep{piga2019performance} wherein BO is used to search for the best (parametrized) prediction model for MPC by directly optimizing closed-loop performance determined from experimental (or high-fidelity simulation) data. We  take a similar perspective in this work and show how the framework is applicable to more general MPC formulations (including economic, nonlinear, robust, and/or output feedback). 

The main focus of this paper is on integrated performance-driven model learning and MPC tuning under uncertainty. The explicit incorporation of uncertainty into the formulation of the automated tuning optimization problem (through both the objective and constraints) is one of the main contributions of this work. In particular, we show how independent uncertainty samples can be used to obtain effectively noisy measurements of objective and constraint, which can be accounted for in the BO method through proper estimation of the noise variance. The second contribution is to leverage a \textit{constrained} variant of BO \citep{gardner2014bayesian,JMLR:v17:15-616} to directly handle output constraints. Constrained BO has been shown to overcome challenges with barrier methods (that penalize constraint violation in the objective) since the constraints are separately modeled and thus provide an independent representation of the feasibility region. 

%performance-driven model learning, there are two key technical contributions presented in this paper. First, we directly account for probabilistic uncertainty

%Although it is well-known the choice of the internal model in MPC has a large influence on the quality of the optimized control actions, model identification is traditionally separated from controller design, even in most automated tuning procedures. 

% maybe cite our scenario method (https://arxiv.org/pdf/2011.07445.pdf) and/or IDC

%In this work, we address the problem of performance-oriented learning of dynamical systems for MPC \citep{bansal2017goal,piga2019performance}. We propose an MPC tuning framework for uncertain nonlinear systems that aims at learning a performance-oriented prediction model for control directly from high-fidelity or experimental data, along with other MPC tuning parameters. The key contribution of this work stems from leveraging constrained BO (CBO) \citep{gardner2014bayesian,JMLR:v17:15-616} to pose a constrained version of the MPC tuning problem that explicitly considers not only the control performance but also system constraints and uncertainties in learning a suitable dynamic model for MPC. We also discuss how the proposed framework can be extended for tuning of output-feedback MPC problems.

%%%%%%%%%%%%%%%%%%%%%%%%%%%%%%
\section{Problem Formulation}
\label{sec:problem-formulation}

Consider an uncertain, time-invariant nonlinear system of the following form
\begin{subequations} \label{eq:sys}
\begin{align}
x_{k+1} &= f_\text{plant}(x_k, u_k, w_k), \\
y_k &= h_\text{plant}(x_k, v_k), 
\end{align}
\end{subequations}
where $k \in \mathbb{N}_0$ denotes the discrete time index, $x_k \in \mathbb{R}^{n_x}$ denotes the system state, $u_k \in \mathbb{R}^{n_u}$ denotes the control input, $y_k \in \mathbb{R}^{n_y}$ denotes the measured output, $w_k \in \mathbb{R}^{n_w}$ denotes the process noise, and $v_k \in \mathbb{R}^{n_v}$ denotes the measurement noise. 
%The functions $f_\text{plant} : \mathbb{R}^{n_x} \times \mathbb{R}^{n_u} \times \mathbb{R}^{n_w} \to \mathbb{R}^{n_x}$ and $h_\text{plant} : \mathbb{R}^{n_x} \times \mathbb{R}^{n_v} \to \mathbb{R}^{n_y}$ represent the system dynamics and measurement equations, respectively. 
The initial state $x_0$, the process noise sequence $w_{0:T-1}$, and the measurement noise sequence $v_{0:T}$ are assumed to be random variables for some finite horizon $T \in \mathbb{N}$.
Control performance $J(y_{1:T}, u_{1:T})$ is assumed to be some known function of the output and input sequences. Since the output depends on the realization of the uncertainty $\delta = (x_0,w_{0:T-1},v_{0:T})$, our actual objective is to minimize expected performance cost
\begin{align}
\mathbb{E}_\delta \{ J(y_{1:T}, u_{1:T}) \} = \int_\Delta J(y_{1:T}, u_{1:T}) p(\delta)d\delta, 
\end{align}
where $p(\delta)$ denotes the probability density function and $\Delta$ is the support of the random vector $\delta$. The controlled system should also satisfy hard input constraints
\begin{align} \label{eq:input-constraints}
u_{0:T} \in \mathbb{U}_{0:T} = \mathbb{U}_0 \times \cdots \times \mathbb{U}_{T},
\end{align}
and joint chance constraints on the output
\begin{align} \label{eq:chance-constraints}
\mathbb{P}_\delta \{  y_{1:T} \in \mathbb{Y}_{1:T} = \mathbb{Y}_1 \times \cdots \times \mathbb{Y}_T \} \geq 1 - \epsilon,
\end{align}
where $\epsilon \in [0,1]$ denotes the allowed violation probability. 
%Note that for simplicity of presentation the constraints \eqref{eq:chance-constraints} are enforced jointly in time, though these can be straightforwardly replaced with separated constraints of the form $\mathbb{P}_\delta \{  y_k  \in \mathbb{Y}_k \} \geq 1 - \epsilon_k$ for all $k = 1,\ldots,T$. 
The ideal controller design problem can be formulated in terms of the following stochastic optimization problem:
\begin{subequations} \label{eq:exact-control-design}
\begin{align}
\min_{\pi \in \Pi} &~~ \mathbb{E}_\delta \{ J(y_{1:T}, u_{1:T}) \}, & \\
\text{s.t.} &~~ x_{k+1} = f_\text{plant}(x_k, u_k, w_k), &\forall k \in \mathbb{N}_0^{T-1}, \\
&~~ y_k = h_\text{plant}(x_k, v_k), &\forall k \in \mathbb{N}_0^{T}, \\
&~~ u_k = \pi_k(y_{0:k}) \in \mathbb{U}_k, &\forall k \in \mathbb{N}_0^{T}, \\
&~~\mathbb{P}_\delta \{  y_{1:T} \in \mathbb{Y}_{1:T} \} \geq 1 - \epsilon,
\end{align}
\end{subequations}
where $\pi = \{ \pi_0, \pi_1, \ldots, \pi_{T} \}$ denotes the control \textit{policy} that is composed of a sequence of control laws $\pi_k : \mathbb{R}^{n_y (k+1)} \to \mathbb{R}^{n_u}$, which are arbitrary functions of the \textit{history} of measured outputs and $\Pi$ denotes the set of possible control policy structures. This problem is extremely challenging for several reasons including the set of functions $\pi \in \Pi$ is infinite dimensional and the probabilistic operators $\mathbb{E}_\delta \{ \cdot \}$ and $\mathbb{P}_\delta \{ \cdot \}$ cannot be computed exactly for general nonlinear systems. Another key complication considered in this work is that functions $f_\text{plant}(\cdot)$ and $h_\text{plant}(\cdot)$ and uncertainty distribution $\delta \sim p(\delta)$ may not be explicitly known. 
%Two situations this would occur are: (i) we only have access to an experimental system in which we can supply control inputs and measure the corresponding outputs or (ii) we have developed a high-fidelity simulation model 
%(e.g., multiple interacting numerical codes/algorithms) 
%such that expressions for $f_\text{plant}(\cdot)$ and $h_\text{plant}(\cdot)$ are not available. 
Thus, \eqref{eq:exact-control-design} belongs to the challenging class of  \textit{constrained black-box optimization under uncertainty} problems. 

Since the optimization problem \eqref{eq:exact-control-design} cannot be solved directly, one must resort to a heuristic controller design strategy
%Traditionally these strategies were a mixture of PID control and logical rules (sometimes referred to as an ``expert system''); however, it has become more common to make control decisions using an advanced control strategy that relies on some underlying optimization problem. 
such as model predictive control (MPC) \citep{rawlings2009model}
%, which is one of the most popular optimization-based methods for controlling constrained, multivariable systems. 
The standard MPC design strategy involves: (i) identification of a control-relevant model of the system \eqref{eq:sys} using first-principles or system identification techniques; (ii) specification of the MPC controller using this control-relevant model to make internal predictions about the future system behavior to select the optimal control inputs; and (iii) trial-and-error experimentation to select the remaining MPC tuning parameters (e.g., prediction horizon, input and output weight matrices, and constraint backoffs) that can have a strong influence on closed-loop performance. Not only are the tuning parameters selected in an ad hoc way in this approach, it is difficult to know how the quality of the predictions of the identified model will translate onto control performance. For example, the I4C methodology mentioned previously is based on the fact that the best model for model-based controller design may not be the one that provides the smallest output prediction errors \citep{gevers2005identification}. 

To address these challenges, we take a \textit{performance-driven} perspective in this work in which the \textit{model} of \eqref{eq:sys} is treated as a degree of freedom that can be used to minimize the closed-loop performance cost. The proposed MPC parametrization is summarized in the next section, which is followed by a description of an automated procedure for simultaneously selecting the internal model and other relevant MPC tuning parameters.

%%%%%%%%%%%%%%%%%%%%%%%%%%%%%%
\section{Policy Approximation: Parametrized Model Predictive Control}
\label{sec:policy-approx}

One of the most critical components of MPC is the model used to predict the outputs given a sequence of inputs. Let the control-relevant model be defined as
\begin{subequations} \label{eq:cr-model}
\begin{align} 
s_{k+1} &= f_\text{mod}(s_k, u_k ; \theta_m), \\
y_k &= h_\text{mod}(s_k ; \theta_m), 
\end{align}
\end{subequations}
where $s_k \in \mathbb{R}^{n_s}$ is the state of the control-relevant model at time step $k \in \mathbb{N}_0$ and $\theta_m \in \mathbb{R}^{n_{\theta,m}}$ is the set of model parameters. %(e.g., for a linear model this would represent a vector composed of the stacked elements of the system matrices). 
Although written in state-space form, \eqref{eq:cr-model} could be a realization of an input-output black-box model such as a nonlinear auto-regressive model with exogenous variables (NARX) in which case $\theta_m$ would represent the coefficients of the selected basis functions.

Given the approximate input-output system model \eqref{eq:cr-model}, the MPC controller solves the following finite-horizon optimal control problem at each sampling time $k \in \mathbb{N}_0^T$:
\begin{subequations} \label{eq:mpc-optimization}
\begin{align} \label{eq:mpc-optimization-cost}
\min_{\mathbf{U}_k} &~~ \textstyle \sum_{i=1}^{N_p} \tilde{\ell}(y_{i|k},u_{i|k} ; \theta_o), & \\
\text{s.t.} &~~ s_{i+1|k} = f_\text{mod}(s_{i|k}, u_{i|k} ; \theta_m), &\forall i \in \mathbb{N}_0^{N_p-1}, \\
&~~ y_{i|k} = h_\text{mod}(s_{i|k} ; \theta_m), &\forall i \in \mathbb{N}_1^{N_p}, \\
&~~ (y_{i|k} + \theta_b, u_{i|k}) \in \mathbb{Y}_{k+i} \times \mathbb{U}_{k+i}, &\forall i \in \mathbb{N}_1^{N_p}, \\
&~~ s_{0|k} = \hat{s}_k,
\end{align}
\end{subequations}
where $N_p$ denotes the prediction horizon; $s_{i|k}$ and $u_{i|k}$ are, respectively, the predicted state and control input at time $k+i$ given information at time $k$; $\mathbf{U}_k = \{ u_{0|k}, \ldots, u_{N-1|k}\}$ is the control input sequence; $\hat{s}_k \in \mathbb{R}^{n_s}$ is the current state estimate; $\tilde{\ell} : \mathbb{R}^{n_y} \times \mathbb{R}^{n_u} \times \mathbb{R}^{n_{\theta,o}} \to  \mathbb{R}$ is the stage cost function parametrized by $\theta_o \in \mathbb{R}^{n_{\theta,o}}$; 
%$g_k : \mathbb{R}^{n_y} \to \mathbb{R}^{n_{\theta,b}}$ is a function that specifies the nonlinear output constraints $\mathbb{Y}_k = \{ y : g_k(y) \leq 0 \}$ at each time $k$, 
and $\theta_b \in \mathbb{R}^{n_{\theta,b}}$ are the constraint backoff parameters that can be used to improve the robustness properties of the controller, as discussed in, e.g., \citep{paulson2018nonlinear}. 
%Let $\theta_c = (\theta_m, \theta_o, \theta_b, N_p)$ be the concatenation of all changeable parameters in the MPC problem. 
The closed-loop performance index $J(y_{1:T}, u_{1:T})$ is not necessarily the same as the MPC cost function \eqref{eq:mpc-optimization-cost}. In most cases $J(y_{1:T}, u_{1:T})$ will be some function that reflects engineering and economic goals, while $\tilde{\ell}(\cdot)$ can be substantially simplified to facilitate online optimization.

Let $\{ u^\star_{i|k}(\hat{s}_k ; \theta_m, \theta_o, \theta_b, N_p) \}_{i=0}^{N_p}$ denote the optimal solution to \eqref{eq:mpc-optimization} for a given state estimate and set of tuning parameters. The receding-horizon control law, implicitly defined in terms of the solution to \eqref{eq:mpc-optimization}, is given by
\begin{align} \label{eq:control-law}
\kappa_c(\hat{s}_k ; \theta_m, \theta_o, \theta_b, N_p) = u^\star_{0|k}(\hat{s}_k ; \theta_m, \theta_o, \theta_b, N_p).
\end{align}
Since the control-relevant state may not be directly measured, we assume a parametrized state estimator of the following form exists:
\begin{align} \label{eq:estimator}
\hat{s}_{k+1} = \kappa_{e}(\hat{s}_k, u_{k}, y_{k+1} ; \theta_e),
\end{align}
where $\theta_e \in \mathbb{R}^{n_{\theta,e}}$ denotes any free parameters in the estimator. For example, $\theta_e$ could represent the process and measurement noise matrices in an extended Kalman filter (EKF) \citep{hoshiya1984structural}.
%or the initial condition of the state estimate. 
By combining the specified controller \eqref{eq:control-law} and estimator \eqref{eq:estimator} structures with the system dynamics \eqref{eq:sys}, we can represent the closed-loop system in terms of the augmented state $z_k = (x_k, \hat{s}_k)$
\begin{subequations} \label{eq:closed-loop-system}
\begin{align}
z_{k+1} &= f_\text{cl}(z_k, w_k; \theta), \\
u_k &= \kappa_\text{cl}(z_k ; \theta), \\
y_k &= h_\text{cl}(z_k, v_k), 
\end{align}
\end{subequations}
where $f_\text{cl} : \mathbb{R}^{n_z} \times \mathbb{R}^{n_w} \times \mathbb{R}^{n_\theta} \to \mathbb{R}^{n_z}$ defines the autonomous closed-loop dynamics, $\kappa_\text{cl} : \mathbb{R}^{n_z} \times \mathbb{R}^{n_\theta} \to \mathbb{R}^{n_u}$ is the control function in the augmented space, $h_\text{cl} : \mathbb{R}^{n_z} \times \mathbb{R}^{n_v} \to \mathbb{R}^{n_y}$ is the measurement function in the augmented space, $\theta = (\theta_m, \theta_o, \theta_b, N_p, \theta_e)$ denotes the complete set of tuning parameters appearing in the control structure, and $n_z = n_x + n_s$ and $n_\theta = n_{\theta,m} + n_{\theta,o} + n_{\theta,b} + n_{\theta,e} + 1$ are the augmented state and tuning parameter dimensions, respectively. Under this restricted class of control policies, we can construct an approximation to \eqref{eq:exact-control-design} as follows:
\begin{subequations} \label{eq:approximate-control-design}
\begin{align}
\min_{\theta \in \Theta} &~~ \mathbb{E}_\delta \{ J(y_{1:T}, u_{1:T}) \}, \\
\text{s.t.} &~~ \text{\eqref{eq:closed-loop-system}},~ \mathbb{P}_\delta \{  y_{1:T} \in \mathbb{Y}_{1:T} \} \geq 1 - \epsilon.
\end{align}
\end{subequations}
Note that the control input constraints \eqref{eq:input-constraints} are not included in this formulation as they are directly enforced by the MPC law \eqref{eq:mpc-optimization}. The key difference between \eqref{eq:approximate-control-design} and \eqref{eq:exact-control-design} is that the proposed approximation \eqref{eq:approximate-control-design} optimizes over a finite dimensional space $\Theta \subset \mathbb{R}^{n_\theta}$. The problem is also a single stage one in which a full simulation can be carried out once the tuning parameters $\theta$ are fixed. As such, we do not have to worry about adapting the control policy over the time steps $k = 1,\ldots,T$, as this is implicitly done through the MPC law defined above. However, this problem is still not immediately solvable due to the presence of the probabilistic operators.
% which must also be approximated because we are unable to exactly propagate the random uncertainties $\delta$ through black-box functions. 
We attempt to address this challenge next using a simulation optimization (SO) paradigm \citep{amaran2016simulation}.
%\begin{align}
%\min_{\theta \in \Theta} &~~ \mathbb{E}_\delta \{ J(y_{1:T}, u_{1:T}) \}, & \\
%\text{s.t.} &~~ \text{closed-loop system \eqref{eq:closed-loop-system}}, \\
%&~~ \mathbb{P}_\delta \{  y_{1:T} \in \mathbb{Y}_{1:T} \} \geq 1 - \epsilon,
%\end{align}

\begin{remark}
It is important to note that the number of model parameters $n_{\theta,m}$ in \eqref{eq:cr-model} can be quite large in generic black-box models, which inherently increases the complexity of the problem. As such, it is advised to encode as much prior knowledge as possible in a particular problem at hand, which will typically result in a gray-box model defined in a reduced parametric space. 
\end{remark}

\begin{remark}
Although we focused on a nominal MPC formulation above, this could straightforwardly be replaced with robust or stochastic MPC methods that directly account for uncertainty within the predictions. However, this will come at the cost of more expensive closed-loop simulations as well as the need to develop an uncertainty description. Again, one can imagine that some parameters of the uncertainty distribution $\theta_u$ could also be treated as controller tuning parameters. 
\end{remark}

\begin{remark}
Whenever the state of the model \eqref{eq:cr-model} is measurable then we can ignore the state estimator \eqref{eq:estimator}. It is worthwhile noting, however, that the estimator could still be useful whenever it is desired to adapt some of the model parameters online due to unexpected drifts or faults. 
\end{remark}

%%%%%%%%%%%%%%%%%%%%%%%%%%%%%%
\section{Proposed Closed-loop Simulation Optimization Method}
\label{sec:closed-loop-opt}

Let $f(\theta) = \mathbb{E}_\delta \{ J(y_{1:T}, u_{1:T}) \}$ denote the expected performance and $c(\theta) = \mathbb{P}_\delta \{  y_{1:T} \in \mathbb{Y}_{1:T} \} -1 + \epsilon$ be a shorthand for the chance constraint. We can now restate \eqref{eq:approximate-control-design} in the following compact manner:
\begin{align} \label{eq:simulation-optimization}
\min_{\theta \in \Theta} ~ f(\theta) ~~ \text{s.t.} ~~ c(\theta) \geq 0.
\end{align}
As discussed in \citep{paulson2018nonlinear}, the uncertainty propagation steps needed to evaluate $f(\theta)$ and $c(\theta)$ can be performed using a variety of different techniques.
%such as Monte Carlo (MC) simulation, Taylor series, unscented transform (UT), or polynomial chaos expansions (PCEs). 
Since the number of uncertainties $n_\delta = n_x + n_w T + n_v(T+1)$ grows with the number of time steps $T$, Monte Carlo (MC) sampling is a likely a good choice since its convergence rate is known to be independent of the $n_\delta$ and instead only depends on the number of samples. Using MC sampling, the objective and constraints in \eqref{eq:simulation-optimization} can then be approximated as \citep{kleywegt2002sample}
\begin{subequations} \label{eq:sample-average-approx}
\begin{align}
f(\theta) &\approx M^{-1} \textstyle\sum_{j=1}^M J(y_{1:T}^j, u_{1:T}^j), \\
c(\theta) &\approx M^{-1} \textstyle\sum_{j=1}^M \big( \mathbf{1}_{\mathbb{Y}_{1:T}}( y_{1:T}^j ) - 1 + \epsilon \big),
\end{align}
\end{subequations}
where $M$ denotes the number of samples, $\mathbf{1}_A(x)$ denotes the indicator function over the set $A$ (1 when $x \in A$ and 0 otherwise), $\{ \delta^1,\ldots,\delta^M \}$ are independent and identically distributed (i.i.d.) samples of $\delta \sim p(\delta)$, and $(y_{1:T}^j, u^j_{1:T})$  are the simulated closed-loop output and input sequences given the $j$th uncertainty sample $\delta^j$. It is worth noting that our only assumption about the system thus far is that we have the ability to execute \eqref{eq:closed-loop-system} for any choice of tuning parameters $\theta \in \Theta$, and every simulation is performed under an i.i.d. sample of $\delta$. 

Since a new set of uncertainty samples are drawn every time we run the closed-loop ``simulator'' \eqref{eq:closed-loop-system}, the sample average approximations in \eqref{eq:sample-average-approx} produce stochastic/noisy observations of the objective and the constraints
\begin{align} \label{eq:observations-noise}
y^f &= f(\theta) + \varepsilon^f, ~~ y^c = c(\theta) + \varepsilon^c,
\end{align}
where $\varepsilon^f$ and $\varepsilon^c$ represent the observation noise for $f(\theta)$ and $c(\theta)$, respectively. For any $M$, these estimators are unbiased, i.e., $\mathbb{E}_{\delta^1,\ldots,\delta^M}\{ y^f \mid \theta \} = f(\theta)$ and $\mathbb{E}_{\delta^1,\ldots,\delta^M}\{ y^c \mid \theta \} = c(\theta)$; however, these observations may have relatively high variance unless $M$ is large. We assume it is not possible to select a large $M$ due to the fact that the closed-loop simulation or experiment is prohibitively costly (either from a computational or monetary cost point-of-view). Instead these functions can only be evaluated on the order of 100 times or less, meaning $M$ cannot be large enough to ignore noise in the observations. As such, we want to utilize an algorithm that can systematically explore the tuning parameter space (relative to random or grid-based search methods) that can also accommodate noisy objective and constraint observations. 

Bayesian optimization (BO) is a family of algorithms that can solve black-box optimization problems in the presence of noisy observations. The basic idea in BO is to construct a surrogate model of the objective function $f(\theta)$ using a set of initial $n$ observations denoted by $\mathcal{D}_n = \{ (\theta_{1:n}, y^f_{1:n}) \}$. The statistical surrogate model thus provides a posterior distribution of the objective function that can be combined with an acquisition function to decide where to sample next. The acquisition function is defined in a way to tradeoff between uncertainty (related to variance) and performance (related to mean) at unexplored points $\theta \in \Theta$. The new observation is added to the data set, i.e., $\mathcal{D}_{n+1} = \{ \mathcal{D}_n, (\theta_{n+1}, y^f_{n+1}) \}$, which is then used to update the surrogate model. This process continues until the surrogate model converges to the global solution or the maximum number of iterations is reached. See, e.g., \cite{shahriari2015taking} for a recent review of BO and \cite{lu2020mpc} for an application to MPC tuning in central heating, ventilation, and air conditioning (HVAC) plants. 

First, we discuss the traditional BO strategy for \eqref{eq:simulation-optimization} that neglects constraints $c(\theta) \geq 0$ and then describe an extension that modifies the acquisition function to account for the probability that constraints are satisfied. In this work, we exclusively use Gaussian process (GP) surrogate models \citep{rasmussen2003gaussian}. In particular, we assume that the objective function has a GP prior of the form $f(\theta) \sim \mathcal{GP}(\mu_0, k)$ where $\mu_0 : \Theta \to \mathbb{R}$ is the prior mean and $k : \Theta \times \Theta \to \mathbb{R}$ is the prior covariance function. There are many possible choices for the covariance function; in this work we selected the squared exponential function
\begin{align}
k(\theta, \theta') = \exp\left( -\frac{1}{2 l^2} \| \theta - \theta' \|^2 \right),
\end{align}
where $l$ is a length-scale parameter. Under the GP prior and assumed i.i.d. Gaussian noise $\varepsilon^f \sim \mathcal{N}(0,\sigma^2)$, the function evaluations $y^f_{1:n}$ are jointly Gaussian:
\begin{align}
y^f_{1:n} \sim \mathcal{N}(\mathbf{m}, \mathbf{K} + \sigma^2 I_n),
\end{align}
where $[\mathbf{m}]_i = \mu_0(\theta_i)$ are the elements of the mean vector and $\mathbf{K} \in \mathbb{R}^{n \times n}$ is a symmetric covariance matrix with elements $[\mathbf{K}]_{i,j} = k(\theta_i,\theta_j)$. This implies the corresponding function value $f(\theta)$ at any test point $\theta$ must be jointly Gaussian with $y^f_{1:n}$, i.e.,
\begin{align}
& \begin{bmatrix}
y^f_{1:n} \\
f(\theta)
\end{bmatrix} \sim \mathcal{N}\left( \begin{bmatrix}
\mathbf{m} \\
\mu_0(\theta)
\end{bmatrix} , \begin{bmatrix}
\mathbf{K} + \sigma^2 I_n &\mathbf{k}(\theta) \\
\mathbf{k}(\theta)^\top &k(\theta,\theta)
\end{bmatrix} \right).
\end{align}
where $\mathbf{k}(\theta) \in \mathbb{R}^n$ is a vector of covariance terms between $\theta_{1:n}$ and $\theta$, i.e., $[\mathbf{k}(\theta)]_i = k(\theta_i,\theta)$. Due to the properties of joint Gaussian random variables, we find that the posterior distribution $p(f(\theta) | y^f_{1:n}, \theta_{1:n}, \theta)$ of the objective given all available noisy observations is Gaussian with the following mean and covariance \citep{rasmussen2003gaussian}:
\begin{subequations} \label{eq:posterior-mean-variance}
\begin{align}
\mu_{n}(\theta) &= \mu_0(\theta) + \mathbf{k}(\theta)^\top (\mathbf{K} + \sigma^2 I_n ) ( y^f_{1:n} - \mathbf{m} ), \\
\sigma^2_{n}(\theta) &= k(\theta,\theta) - \mathbf{k}(\theta)^\top (\mathbf{K} + \sigma^2 I_n)^{-1} \mathbf{k}(\theta).
\end{align}
\end{subequations}
Given this posterior distribution, we still need to optimize over an acquisition function $\alpha(\theta ; \mathcal{D}_n)$ to search for the next optimal sampling point $\theta_{n+1}$. Although several are available, we focus on the expected improvement (EI) criteria that measures the expected amount by which the current objective will be improved over some incumbent (or current best) solution. The EI acquisition function has been shown to be analytically computable for a GP model
\begin{align} \label{eq:expected-improvement}
\alpha_\text{EI}(\theta ; \mathcal{D}_n) = \sigma_n(\theta)(z_n(\theta) \Phi(z_n(\theta)) + \phi(z_n(\theta)),
\end{align}
where $z_n(\theta) = (\eta - \mu_n(\theta)) / \sigma_n(\theta)$, $\Phi$ and $\phi$ are, respectively, the standard Gaussian cumulative density and probability density functions, and $\eta$ is the incumbent solution. The ``best'' choice of $\eta$ depends on the context; for deterministic objective evaluations, it is often set to the best observed value $\eta = \max_{i \in \{ 1,\ldots,n \}} y_i^f$. However, as discussed in \citep{wang2014theoretical}, this choice can be quite fragile in the stochastic setting and thus we use the best mean value $\eta = \min_{\theta \in \Theta} \mu_n(\theta)$ as a more reasonable alternative. 
% https://arxiv.org/pdf/1406.7758.pdf; Theoretical Analysis of Bayesian Optimisation with Unknown Gaussian Process Hyper-Parameters

To handle the black-box constraints, we need to modify the acquisition function to show improvement only when $c(\theta) \geq 0$ holds. Similarly to the objective, we model the constraint function $c(\theta)$ with a GP prior whose evaluations are corrupted with Gaussian noise. We must then weight the original EI in \eqref{eq:expected-improvement} by the probability of the constraints being satisfied. This results in the expected improvement with constraints (EIC) that can be analytically computed as follows \citep{gardner2014bayesian}:
\begin{align} \label{eq:constrained-acquisition}
\alpha_\text{EIC}(\theta ; \mathcal{D}_n) &= \alpha_\text{EI}(\theta ; \mathcal{D}_n) \mathbb{P}\{ c(\theta) \geq 0 | \mathcal{D}_n^c \}, \\\notag
&= \alpha_\text{EI}(\theta ; \mathcal{D}_n) \Phi\left( \frac{\mu_n^c(\theta)}{\sigma_n^c(\theta)} \right),
\end{align}
where $\mathcal{D}_n^c = \{ (\theta_{1:n}, y^c_{1:n}) \}$ is the set of $n$ noisy constraint observations and $\mu^c_n(\theta)$ and $\sigma_n^c(\theta)$ are, respectively, the posterior predictive mean and standard deviation of the GP surrogate model for $c(\theta)$ -- similarly defined to \eqref{eq:posterior-mean-variance} for the objective. Due to the presence of constraints, we must modify the definition of the incumbent value $\eta$ to be the best mean value such that the constraints are satisfied. Since we cannot guarantee the constraints will be satisfied given only noisy observations, it has been suggested to require constraint satisfaction with some probability:
\begin{align} \label{eq:constrained-incumbent}
\eta = \min_{\theta \in \Theta} ~ \mu_n(\theta) ~~ \text{s.t.} ~~ \Phi\left( \frac{\mu_n^c(\theta)}{\sigma_n^c(\theta)} \right) \geq 1 - \beta,
\end{align}
for a relatively small $\beta \in (0,1)$. However, we still cannot guarantee a feasible solution to \eqref{eq:constrained-incumbent} exists due to, e.g., limited data available for the constraints. To overcome this challenge, one can ignore the factor $\alpha_\text{EI}(\theta ; \mathcal{D}_n)$ in \eqref{eq:constrained-acquisition} to search only for a feasible point by maximizing the probability of constraint satisfaction \citep{gelbart2014}. 
%This feasibility search is purely exploitative and will discover that either a particular region is feasible or its probability will drop and it will move onto a more promising region. 

\begin{figure}[tb!]
\centering
\includegraphics[width=0.9\linewidth]{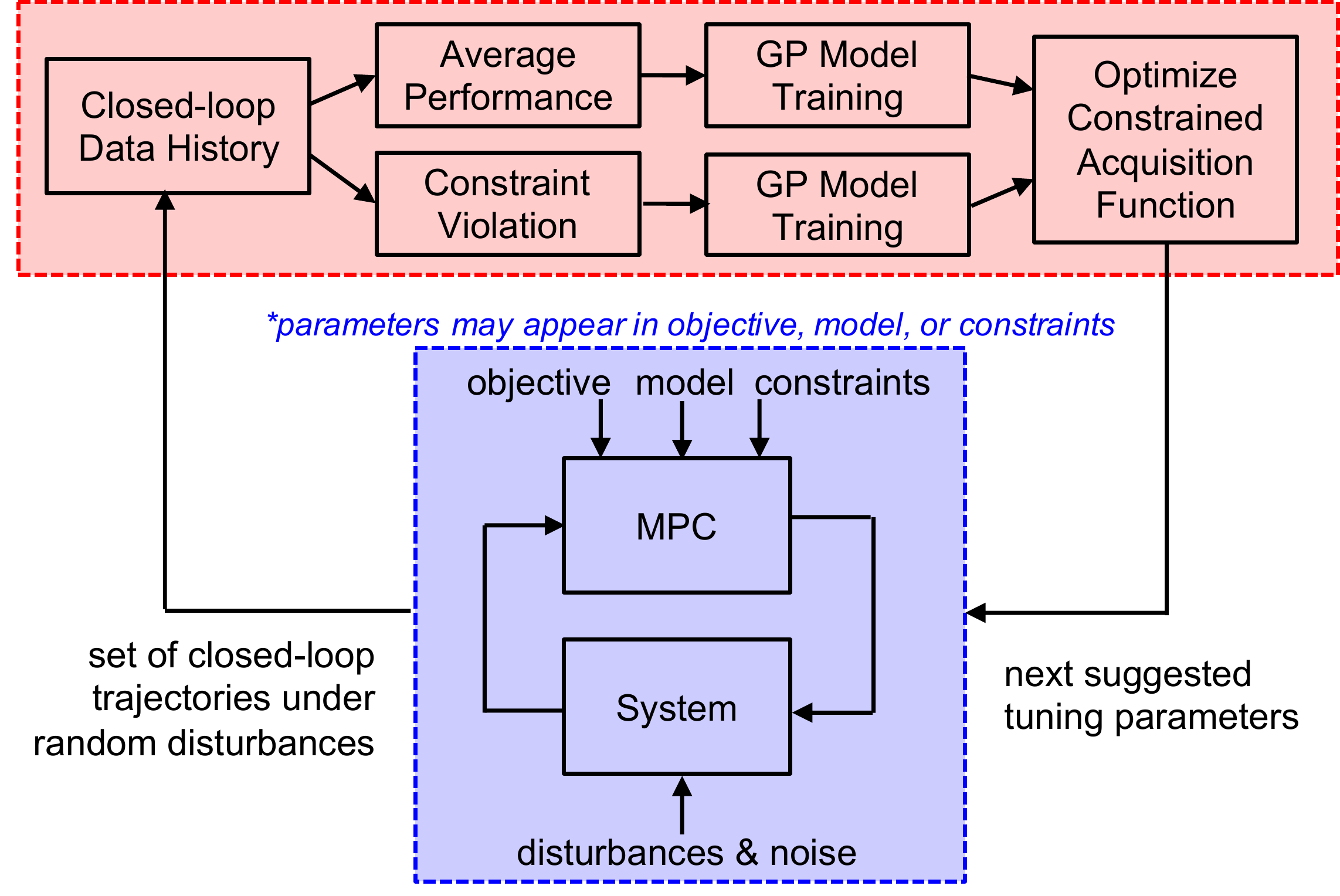}
\caption{Illustration of how constrained BO method can be used for data-driven automatic MPC tuning under constraints. The blue box denotes the closed-loop experiment or simulation that must be performed 1 or more times for fixed tuning parameters. The red box shows the constrained BO steps, which must be executed to decide the next best tuning parameter values to use for closed-loop data collection.}
\label{fig:cbo-flowchat-description}
\end{figure}

In our proposed strategy, the next sampling point $\theta_{n+1}$ is obtained by solving the following optimization problem:
\begin{align}
\theta_{n+1} = \argmax_{\theta \in \Theta} ~ \alpha_\text{EIC}(\theta ; \mathcal{D}_n).
\end{align}
Since the acquisition function is cheap to evaluate (relative to $f$ and $c$), its maximization can be carried out more efficiently. An overview of the various methods used to solve this optimization that can avoid local solutions is provided in \citep{gardner2014bayesian}. An illustration of the proposed automated MPC tuning strategy under uncertainty using constrained BO is shown in Figure \ref{fig:cbo-flowchat-description}.

\begin{remark}
There are several hyperparameters in the GP regression models that may appear in the mean function, covariance kernel (e.g., $l$), or the observation model (e.g., noise level $\sigma^2$). There are two main ways for handling these hyperparameters. The most involved way is to marginalize out the uncertainty in these hyperparameters by determining their posterior distribution given the data $\mathcal{D}_n$ using Bayes' rule. The acquisition function must then be integrated over this posterior, which can be approximated using some sampling strategy. A simpler approach is to approximate this integral in terms of a Dirac delta measure for the posterior at the maximum likelihood or maximum \textit{a posteriori} point estimate. These point estimates can be found by either maximizing the marginal likelihood or the unnormalized posterior for which standard optimization methods are readily available. 
\end{remark}

\begin{remark}
The assumed Gaussian noise model with constant variance for the objective and constraint evaluations in \eqref{eq:observations-noise} is an approximation made for simplicity. In reality, the noise level may depend on the specific value of $\theta$, which should be accounted for to obtain a more accurate GP surrogate model. One way to do this is to model the noise with a second GP model, as discussed in \citep{goldberg1998regression}. Future work is needed to best incorporate input-dependent noise models into constrained BO. 
\end{remark}
% Regression with Input-dependent Noise: A Gaussian Process Treatment
% http://papers.nips.cc/paper/1444-regression-with-input-dependent-noise-a-gaussian-process-treatment.pdf

%%%%%%%%%%%%%%%%%%%%%%%%%%%%%%
\section{Numerical Example}
\label{sec:example}

\subsection{Black-box plant model description}

We demonstrate the proposed approach on a nonlinear continuously stirred tank reactor (CSTR) problem adapted from \citep{paulson2018nonlinear}. The following set of reactions are assumed to take place in the CSTR
\begin{align*}
\text{A} &\to \text{B} \to \text{C}, ~~2\text{A} \to \text{D}.
\end{align*}
The plant model \eqref{eq:sys} is assumed to be a discretized version of the following set of ordinary differential equations
\begin{subequations} \label{eq:ode-cstr}
\begin{align}
\dot{c}_\mathrm{A} &= F(c_{\mathrm{A}0} - c_\mathrm{A}) - k_1 c_\mathrm{A} - k_3 c_\mathrm{A}^2, \\
\dot{c}_\mathrm{B} &= -Fc_\mathrm{B} + k_1 c_\mathrm{A} - k_2 c_\mathrm{B}, \\
\dot{T}_\mathrm{R} &= F(T_\text{in} - T_\mathrm{R}) + \frac{k_\mathrm{W}A}{\rho c_\mathrm{p} V_\mathrm{R}}(T_\mathrm{K} - T_\mathrm{R}) \\\notag
& ~~~ - \frac{k_1c_\mathrm{A}\Delta H_{\mathrm{AB}} + k_2c_\mathrm{B}\Delta H_{\mathrm{BC}} + k_3c_\mathrm{A}^2\Delta H_{\mathrm{AD}}}{\rho c_\mathrm{p}} , \\
\dot{T}_\mathrm{K} &= \frac{1}{m_\mathrm{K}c_{\mathrm{pK}}}(\dot{Q}_\mathrm{K} + k_\mathrm{W} A (T_\mathrm{R} - T_\mathrm{K})).
\end{align}
\end{subequations}
where $c_\mathrm{A}$ and $c_\mathrm{B}$ denote the concentrations of species A and B, respectively, $T_\mathrm{R}$ is the reactor temperature, and $T_\mathrm{K}$ is the coolant temperature. The control input is the feed flowrate $F = \dot{V}_{in}/V_R$ and the reaction kinetics exhibit the following Arrhenius dependence:
\begin{align} \label{eq:rate-cstr}
k_{i}= k_{0,i} e^{\frac{-E_{A, i}}{R\left(T_{R}+273.15\right)}},~~\forall i \in \{1,2,3 \}.
\end{align}
The sampling time was chosen to be $\delta t = 0.005$ h. We assume that only species B and the reactor temperature are measurable online such that $y_k = (c_{\mathrm{B},k}, T_{\mathrm{R},k}) + v_k$ where $v_k \sim \mathcal{N}(0, \Sigma_v)$ is noise, with $\Sigma_v = \text{diag}(\sigma_B^2, \sigma_R^2)$. For simplicity, we did not consider process noise. The model parameters can be found in Table \ref{tab:parameters_CSTR} and the initial conditions and relevant constraints are summarized in Table \ref{tab:initialconditionsconstraints}. Since only noisy temperature measurements are available, the bounds on $T_\mathrm{R}$ are enforced as a chance constraint. Instead of the single constraint \eqref{eq:chance-constraints}, we enforce constraints separately at each time point, i.e., $\mathbb{P}_\delta \{  y_k  \in \mathbb{Y}_k \} \geq 1 - \epsilon_k$ where $\epsilon_k = 0.05$ for all $k = 1,\ldots,T$. This does not change the derivations shown in Section \ref{sec:closed-loop-opt} because we can always define $c(\theta) = \min_{k \in \{ 1,\ldots,T \}} \{ \mathbb{P}_\delta \{  y_k  \in \mathbb{Y}_k \}  -1 + \epsilon_k \} \geq 0$ in terms of the smallest probability over time. We assume that our closed-loop objective is to maximize the production of component B 
\begin{align}
n_{B,Total} = \int_0^{t_f} \dot{n}_B(t)dt \approx \sum_{k=0}^{T-1} V_\text{in} c_{\mathrm{B},k} F_k \delta t,
\end{align}
such that the performance cost is $J(y_{1:T}, u_{1:T}) = -n_{B,Total}$ where $t_f = 0.2$ h is the final time of the experiment and $T = 40$ sampling time intervals. 

\begin{table}[tb]
\caption{Parameter values for CSTR case study.}
\label{tab:parameters_CSTR}
\centering
\begin{tabular}{lll} \hline
\textbf{Parameter} &\textbf{Value} &\textbf{Unit} \\\hline
$k_{0,1}$		& $1.287 ~ 10^{12}$		& \text{h$^{-1}$} \\
$k_{0,2}$		& $1.287 ~ 10^{12}$		& \text{h$^{-1}$} \\
$k_{0,3}$		& $9.043 ~ 10^{9}$		& \text{L~\text{mol}$^{-1}$~\text{h}$^{-1}$} \\
$E_{A,1}/R$		& $9758.3$		& \text{K} \\
$E_{A,2}/R$		& $9758.3$		& \text{K} \\
$E_{A,3}/R$		& $7704.0$		& \text{K} \\
$\Delta H_\mathrm{AB}$		& $4.2$		& \text{kJ~\text{mol}$^{-1}$} \\
$\Delta H_\mathrm{BC}$		& $-11.0$		& \text{kJ~\text{mol}$^{-1}$} \\
$\Delta H_\mathrm{AD}$		& $-41.85$		& \text{kJ~\text{mol}$^{-1}$} \\
$\rho$		& $0.9342$		& \text{kg~\text{L}$^{-1}$} \\
$c_\mathrm{p}$		& $3.01$		& \text{kJ~\text{kg}$^{-1}$~\text{K}$^{-1}$} \\
$c_\mathrm{pK}$		& $2.0$		& \text{kJ~\text{kg}$^{-1}$~\text{K}$^{-1}$} \\
$A$		& $0.215$		& \text{m$^{2}$} \\
$V_\mathrm{R}$		& $10.01$		& \text{L} \\
$m_\mathrm{k}$		& $5.0$		& \text{kg} \\
$T_\mathrm{in}$		& $130.0$		& $^\circ \text{C}$ \\
$k_\mathrm{W}$		& $4032$		& \text{kJ~h$^{-1}$~m$^{-2}$~K$^{-1}$} \\
$\dot{Q}_\mathrm{K}$		& $-4500$		& \text{kJ~h$^{-1}$} \\ \hline
\end{tabular}
\end{table}

\begin{table}[tb]
\centering
\caption{Initial conditions and constraints for CSTR.}
\begin{tabular}{ccccc} \hline
\textbf{Variable} & \textbf{Init. cond.} & \textbf{Min.} &\textbf{Max.} & \textbf{Unit} \\\hline
$c_\text{A}$ &1.0			&-- 	&-- &mol/L \\
$c_\text{B}$ &1.0 			&-- 	&-- &mol/L \\
$T_\text{R}$ &100.0	&100.0 	&150.0 &$^\circ\text{C}$ \\
$T_\text{K}$ &100.0 	&--  &-- &$^\circ\text{C}$ \\
$F$ &-- &5.0 &35.0 &h$^{-1}$ \\ \hline
\end{tabular} \label{tab:initialconditionsconstraints}
\end{table}

% could add citation to table 1 in my IFAC NMPC paper. 

\subsection{Parametrized formulation of MPC policy}

We emphasize the fact that the plant model \eqref{eq:ode-cstr}--\eqref{eq:rate-cstr} is a black-box in this work and so cannot be used in model-based control design. Instead, we look to learn a control-relevant model of the form \eqref{eq:cr-model}. In particular, we focus on an NARX model structure that can be stated as follows
\begin{align} \label{eq:narx}
y_{k} = f_\text{NARX}(y_{k-1},,\ldots,y_{k-L_y},u_{k-1},\ldots,u_{k-L_u}),
\end{align}
where $L_y$ and $L_u$ denotes the number of output and input lags, respectively. The function $f_\text{NARX}$ can be any nonlinear function such as a polynomial or a neural network. In this work, we selected a second-order polynomial that ignores the interaction terms. Although in principle $L_y$ and $L_u$ could be treated as model parameters $\theta_m$, we decided to fix them to $L_y = L_u = 1$ to ensure the tuning parameter space was not too large. This implies that $\theta_m$ is composed of 14 coefficients -- 7 for each $n_y=2$ output that multiplies the basis set $\{ 1, [y_k]_1, [y_k]_2, u_k, [y_k]_1^2, [y_k]_2^2, u_k^2 \}$. We fixed the MPC prediction horizon to be $N_p = 10$, as performance was found to be relatively insensitive to this choice. We selected an economic MPC objective function related to the moles of B produced, i.e., $\tilde{\ell}(y_{i|k},u_{i|k}) = V_\text{in} c_{\mathrm{B},i|k} F_{i|k} \delta t$.
%with a small quadratic penalty on input usage where $r = {\color{red}0.1}$. 
The reactor temperature constraint is active in this problem such that it can easily be violated whenever the prediction model is not very accurate. To address this, we include a backoff parameter $\theta_b$ that we wish to select using the proposed constrained BO method, which results in a total of $n_\theta = 14 + 1 = 15$ tuning parameters to be optimized. 

\subsection{Automated MPC tuning using constrained BO and comparison to sequential model identification}

Given the black-box plant description and the parametrized MPC policy, we can now solve \eqref{eq:simulation-optimization} using the constrained BO method summarized in Section \ref{sec:closed-loop-opt}; the \texttt{bayesopt} function in Matlab can be used to handle the GP construction of the objective and constraints as well as specify the EIC criteria using the ``coupled constraint'' option. We explicitly set the objective and constraint evaluations to being stochastic such that the variance is included as a hyperparameter in the corresponding GP models. We budgeted a total of $N_\text{iter} = 40$ iterations, with the first 5 being randomly chosen within the assumed parameter space $\Theta = [-2,2]^{14} \times [0,0.2]$, i.e., all NARX parameters are bounded by $[-2.2]$ and the backoff $\theta_b \in [0,0.2]$\footnote{The NARX model is constructed using scaled data, so that all inputs and outputs are bounded within $[0,1]$.}. 

\begin{figure}[ht!]
\centering
\includegraphics[width=0.8\linewidth]{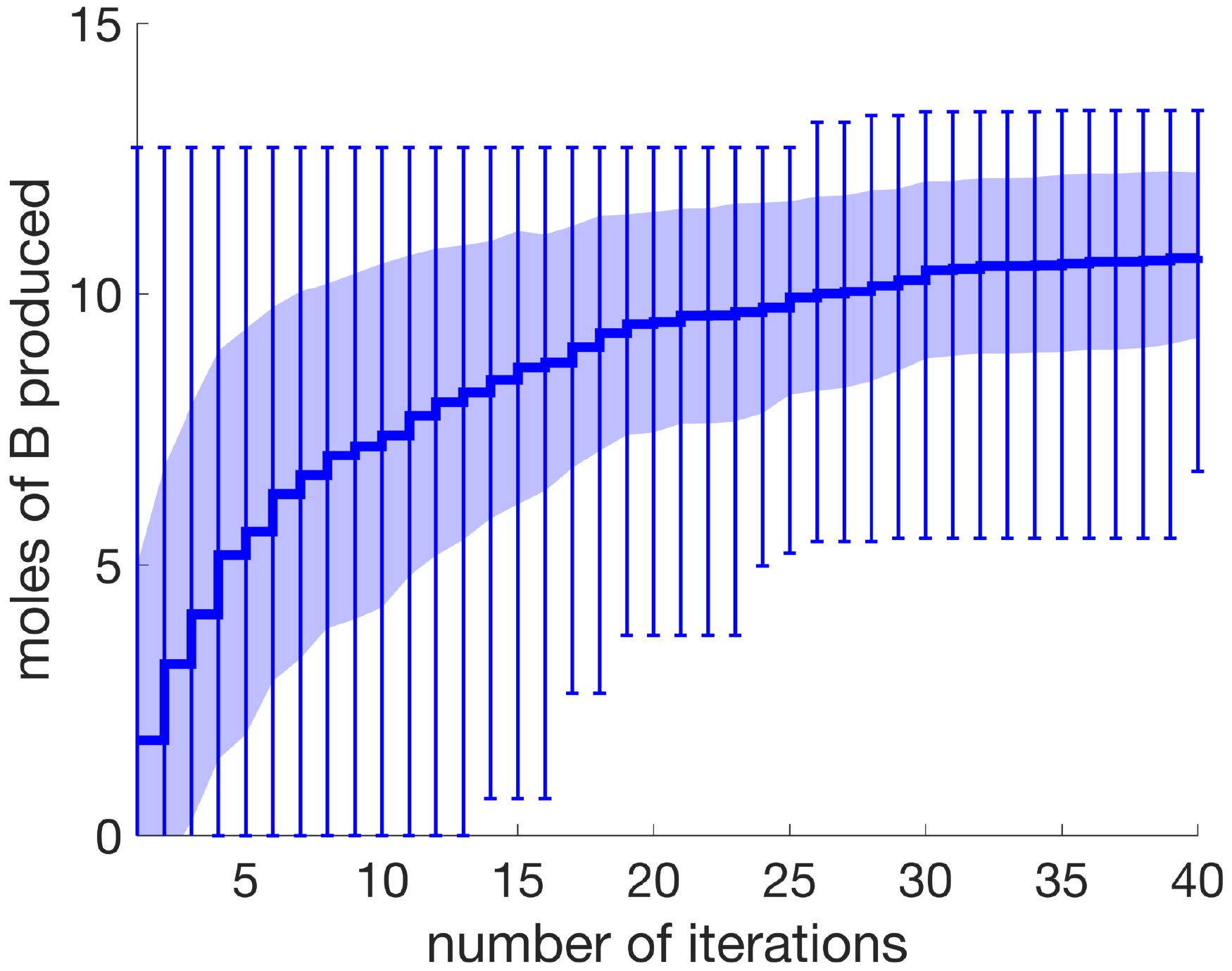}
\caption{The maximum performance (moles of B) versus number of BO iterations for 100 replicate runs. The dark blue line represents the mean value, the shaded region is +/- one standard deviation, and the error bars show the minimum and maximum values. }
\label{fig:bo-iterations}
\end{figure}

\begin{figure}[tb!]
\centering
\begin{subfigure}{0.9\linewidth} % width of left subfigure
\includegraphics[width=0.9\linewidth]{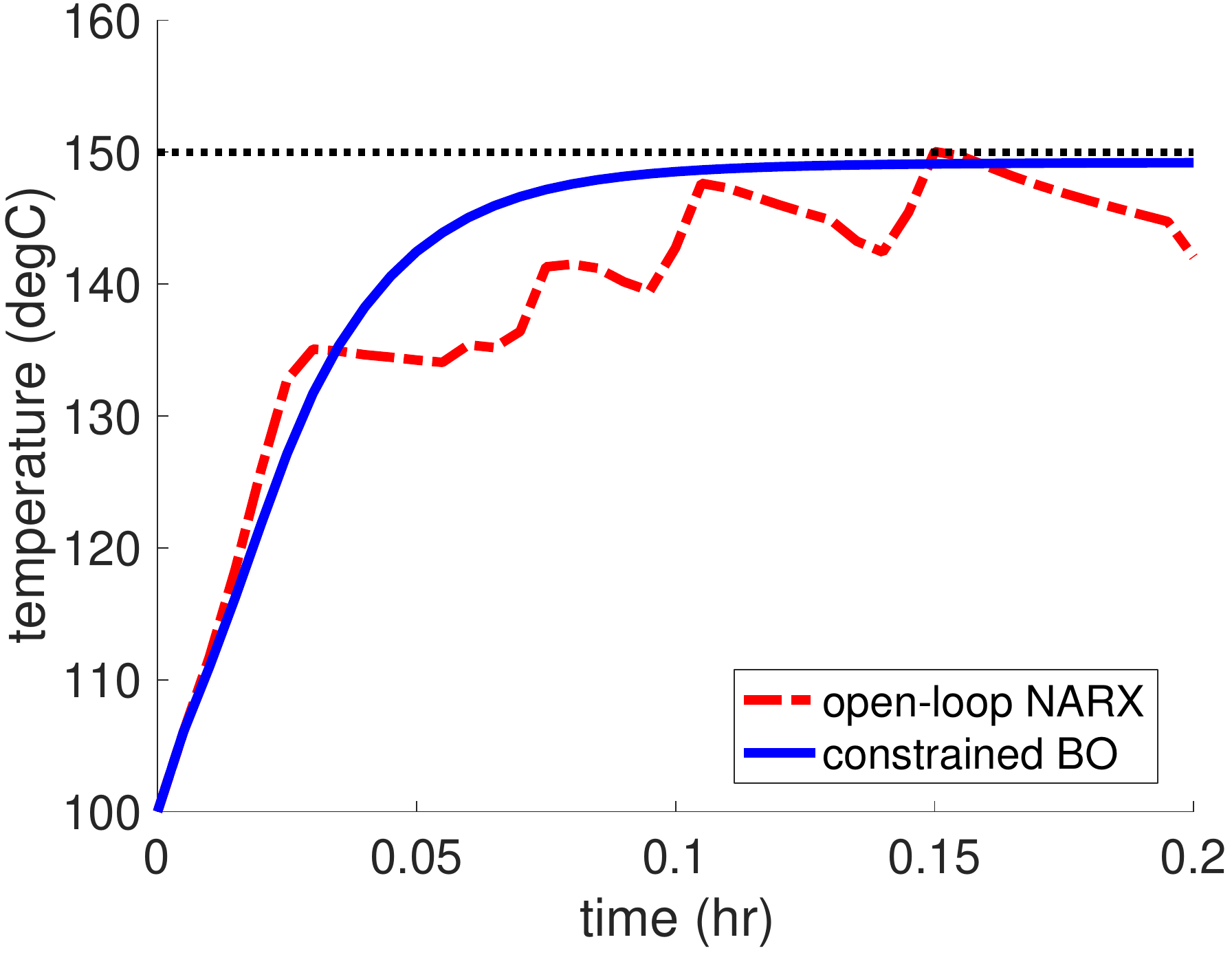}
\caption{temperature profile} % subcaption
\end{subfigure}
\vspace{1em} % here you can insert horizontal or vertical space
\begin{subfigure}{0.9\linewidth} % width of right subfigure
\includegraphics[width=0.9\linewidth]{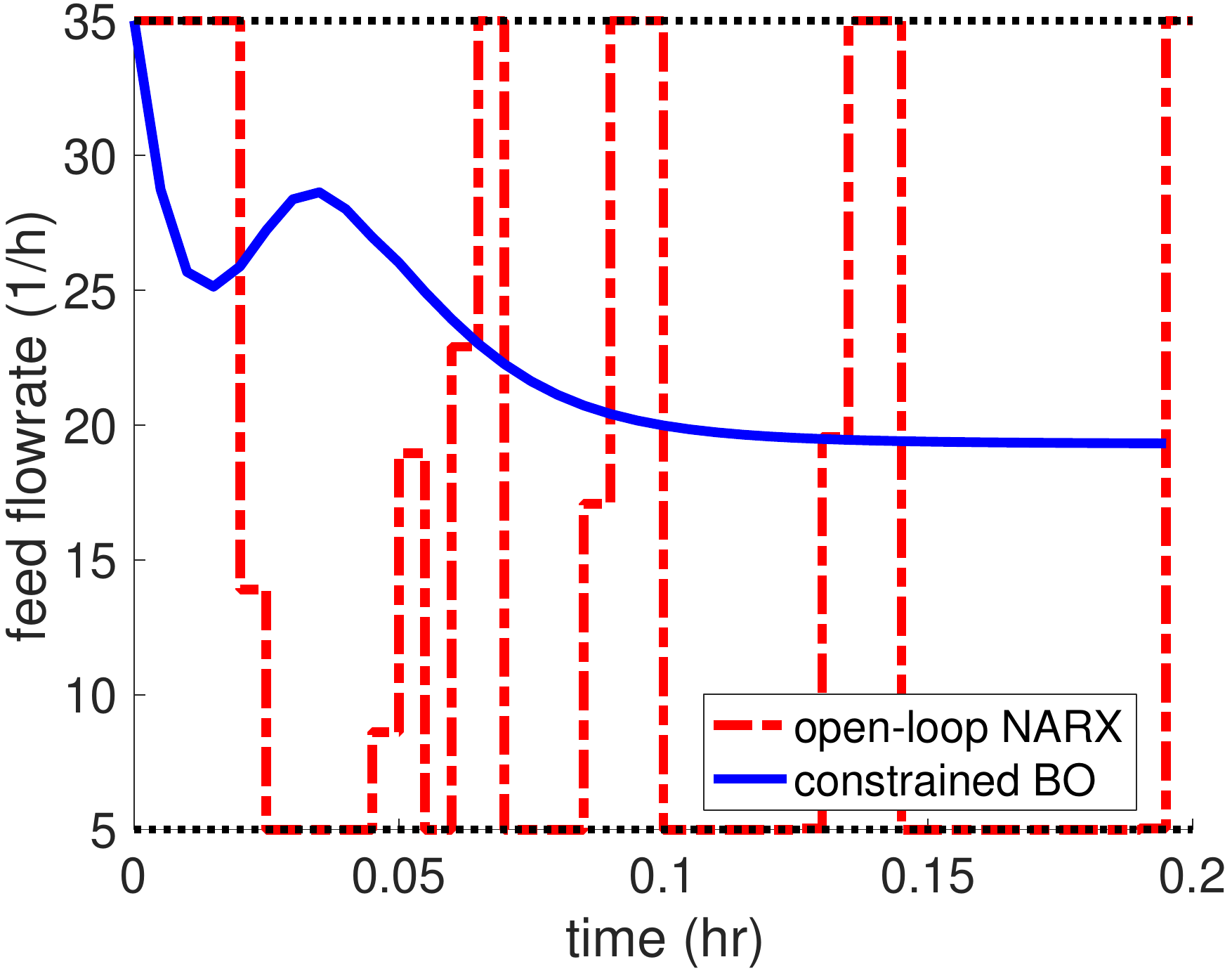}
\caption{feed flowrate profile} % subcaption
\end{subfigure}
\caption{Closed-loop output and input profiles for the best solution obtained with BO compared to an open-loop identified NARX model for the noise-free case.} % caption for whole figure
\label{fig:nominal-results}
\end{figure}

To focus on constraint handling first, we initially set the measurement noise variances to zero. The performance evolution, measured in terms of total moles of B produced, over the number of iterations is shown in Figure \ref{fig:bo-iterations} for 100 independent runs of the constrained BO algorithm. We see a consistent improvement in performance up until around 30 iterations -- we also see that the variance in the final result reduces as number of iterations increases. The resulting closed-loop temperature and feed flowrate profiles are shown in Figure \ref{fig:nominal-results}, which tightly satisfy constraints. To better contextualize these results, we compare this automated model learning procedure to the traditional open-loop NARX identification paradigm (see red dash-dotted lines in Figure \ref{fig:nominal-results}). In particular, we performed pseudo random binary step tests on the plant every 100 time steps to collect a total of 3000 input-output data points. This open-loop data was used to train an NARX model \eqref{eq:narx} using Matlab's System Identification Toolbox, which achieved high accuracy on a holdout set of data, i.e., prediction accuracy of 91.5\%, which is equal to $1-(\text{normalized root mean squared error})$. 

Even though the prediction accuracy is high, the resulting closed-loop performance is poor in the sense that we observe significant variability in the control input profile and minor temperature constraint violations. In fact, the moles of B produced was 5.62, which is less than half of the 12.17 obtained with the constrained BO method. This is likely due to the fact that the input-output data was not sufficiently informative in the regime of high-performance control. As such, we can interpret the proposed strategy as a way to select control-oriented prediction models that are suited to the task at hand -- this is important because the desired set of closed-loop trajectories usually represents a much smaller slice of the input-output space than that used in open-loop identification techniques. 

\begin{figure}[ht!]
\centering
\includegraphics[width=0.8\linewidth]{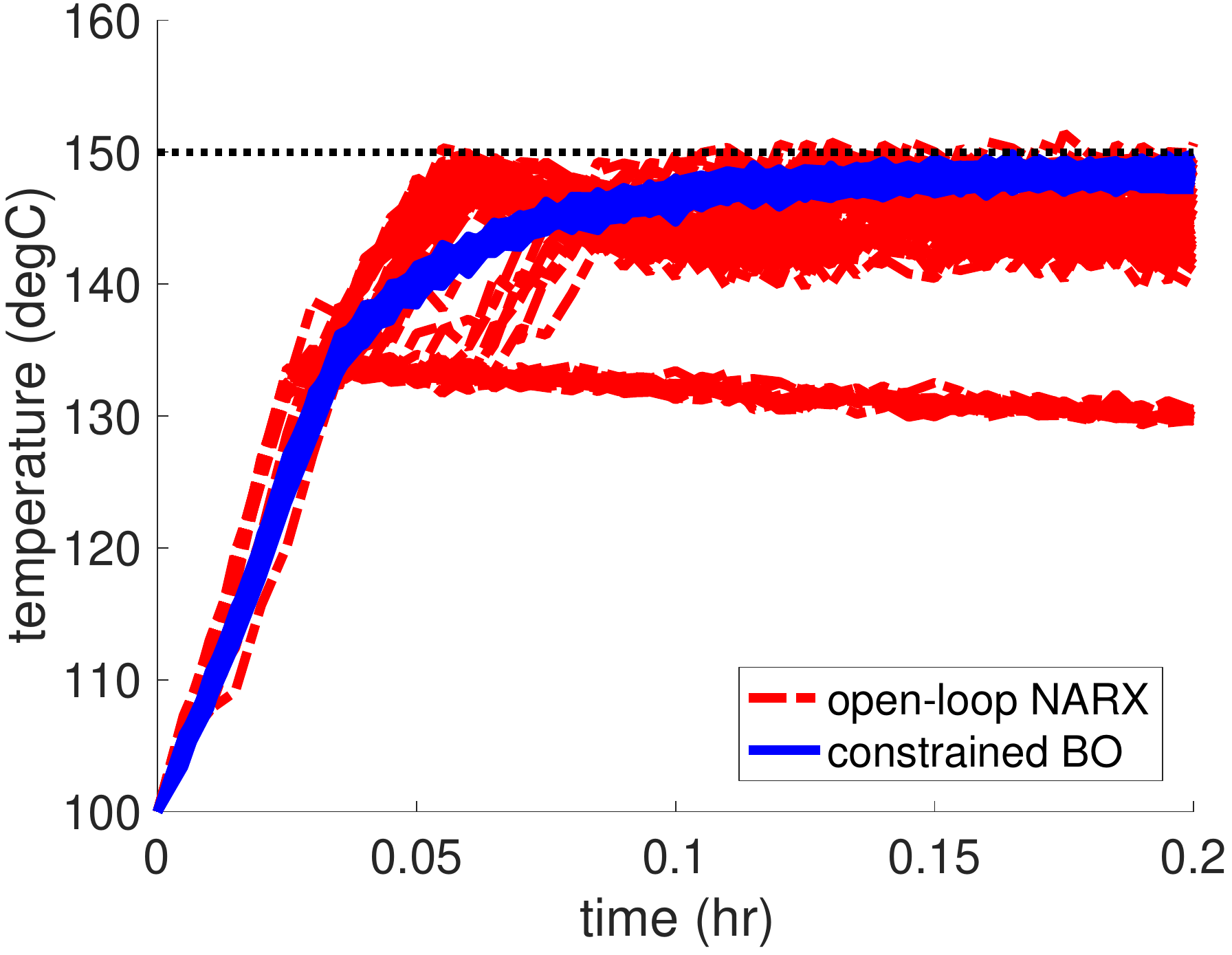}
\caption{Closed-loop temperature profile obtained using the proposed constrained BO method compared to an open-loop identified NARX model in the presence of measurement noise for 100 Monte Carlo realizations.}
\label{fig:stochastic-results}
\end{figure}

Lastly, to demonstrate that the approach is capable of operating in the presence of uncertainty, we repeated the analysis above with measurement noise variances $\sigma_B = 0.2$ mol and $\sigma_R = 10^\circ\text{C}$. To keep a small experimental budget, we selected $M=1$. Due to space limitations, we only show the resulting closed-loop temperature profiles for the open-loop NARX and constrained BO identified models in Figure \ref{fig:stochastic-results}. It is interesting to note that the open-loop identified NARX model appears to be quite susceptible to overfitting, which leads to much higher variability and constraint violations than in the noise-free case. On the other hand, the constrained BO method is able to select NARX coefficients that produce temperature profiles with significantly lower variance and, through proper tuning of the backoff, completely avoids constraint violations. We also emphasize that these results were obtained using the same number of runs as in the deterministic case (only 1 experiment performed for a random noise sequence). This highlights the value of using a GP model that can explicitly account for noisy objective and constraint violations, as it helps guide the search in a way that is not overly optimistic about any one observation. 

%Show BO plot and describe strategy for initializing smarter [start with some open loop data and then improve upon that]. Should mention that we need less total data than OL. We suspect this trend to be true in many problems, though more experimentation is needed to verify this claim. 

%\subsection{Comparison with sequential model identification}
%
%Compare whenever they are given the same amount of data. So basically say (100 runs given we do 100 random open loop experiments and 100 BO closed loop) and see which one better. 

%such that the performance cost $J(y_{1:T}, u_{1:T})$ can be written as
%\begin{align}
%
%\end{align}

%Note that for simplicity of presentation the constraints \eqref{eq:chance-constraints} are enforced jointly in time, though these can be straightforwardly replaced with separated constraints of the form $\mathbb{P}_\delta \{  y_k  \in \mathbb{Y}_k \} \geq 1 - \epsilon_k$ for all $k = 1,\ldots,T$. 

%however, it should be noted the plant model  is treated as a black-box. 
%that we do not assume that we have access to the full plant model \eqref{eq:ode-cstr}--\eqref{eq:rate-cstr} and can only probe the outputs for a given set of control inputs. 

%%%%%%%%%%%%%%%%%%%%%%%%%%%%%%
\section{Conclusions and Future Work}
\label{sec:conclusions}

We present an automated performance-driven MPC tuning strategy for black-box systems in the presence of uncertainty. Since the MPC tuning parameters (which includes the prediction model itself in our approach) affect the closed-loop performance and constraints in complex (generally non-convex and non-smooth) ways, we rely on derivative-free simulation optimization methods to automate the overall tuning process. In particular, we rely on a constrained variant of Bayesian optimization (BO) that utilizes state-of-the-art surrogate models for the objective and constraint functions constructed from noisy observations. By combining these surrogates with a so-called acquisition function, we can iteratively determine the best tuning parameter combination to use for our next closed-loop experiment that directly tradeoffs exploitation of the current best tuning parameters with exploitation of the feasible tuning parameter space (while also accounting for the probability of constraint satisfaction). We demonstrated the effectiveness of the approach on a nonlinear continuously stirred tank reactor case study wherein we observed a more than two-fold improvement in performance compared to the traditional open-loop model identification methods. There are several interesting directions for future work including development of novel constrained BO methods that can more readily scale to high-dimensional problems and that can account for noise levels in the function evaluations that depend on the specific tuning parameter values. 

%This work shows that CBO can be implemented as a useful framework for automating tuning of MPC controllers, even in the presence of uncertainty. By using state-of-the-art surrogate modeling methods, CBO can easily handle noisy objective and constraint values (and general, black-box functions). We reformulated a typical MPC tuning problem into an optimization problem by defining a prediction model and tuning parameters as optimization variables. CBO efficiently solved this optimization problem, wherein a modified acquisition function allowed for consideration of probabilistic constraints, in addition to performance.  Our approach of applying performance-driven automated tuning yielded a two-fold improvement compared to standard, open-loop system identification techniques when evaluated based on a benchmark CSTR. Future work should consider strategies to more efficiently handle high dimensional problems.

% The main point of this work is to show that CBO can be used as a useful paradigm for automating tuning of MPC controllers, even in the presence of uncertainty. It can easily handle noisy objective and constraint values (and general black box functions) by using state-of-the-art surrogate modeling methods. A particularly interesting part of this work is that the model is used as a parameter in the tuning problem. We presented a case study to back this up.

%Maybe remark on high dimensional problems and that being important in future work (could also be in conclusions). 

\bibliography{bibliography}

\begin{thebibliography}{22}
\providecommand{\natexlab}[1]{#1}
\providecommand{\url}[1]{\texttt{#1}}
\providecommand{\urlprefix}{URL }
\expandafter\ifx\csname urlstyle\endcsname\relax
  \providecommand{\doi}[1]{doi:\discretionary{}{}{}#1}\else
  \providecommand{\doi}{doi:\discretionary{}{}{}\begingroup
  \urlstyle{rm}\Url}\fi

\bibitem[{Amaran et~al.(2016)Amaran, Sahinidis, Sharda, and
  Bury}]{amaran2016simulation}
Amaran, S., Sahinidis, N.V., Sharda, B., and Bury, S.J. (2016).
\newblock Simulation optimization: A review of algorithms and applications.
\newblock \emph{Annals of Operations Research}, 240, 351--380.

\bibitem[{Bansal et~al.(2017)Bansal, Calandra, Xiao, Levine, and
  Tomiin}]{bansal2017goal}
Bansal, S., Calandra, R., Xiao, T., Levine, S., and Tomiin, C.J. (2017).
\newblock Goal-driven dynamics learning via {B}ayesian optimization.
\newblock In \emph{Proceedings of the IEEE 56th Annual Conference on Decision
  and Control}, 5168--5173. Melbourne.

\bibitem[{Berkenkamp et~al.(2016)Berkenkamp, Schoellig, and
  Krause}]{berkenkamp2016safe}
Berkenkamp, F., Schoellig, A.P., and Krause, A. (2016).
\newblock Safe controller optimization for quadrotors with {G}aussian
  processes.
\newblock In \emph{Proceedings of the IEEE International Conference on Robotics
  and Automation}, 491--496. Stockholm.

\bibitem[{Forgione et~al.(2019)Forgione, Piga, and
  Bemporad}]{forgione2019efficient}
Forgione, M., Piga, D., and Bemporad, A. (2019).
\newblock Efficient calibration of embedded {MPC}.
\newblock \emph{arXiv preprint arXiv:1911.13021}.

\bibitem[{Gardner et~al.(2014)Gardner, Kusner, Xu, Weinberger, and
  Cunningham}]{gardner2014bayesian}
Gardner, J.R., Kusner, M.J., Xu, Z.E., Weinberger, K.Q., and Cunningham, J.P.
  (2014).
\newblock {B}ayesian optimization with inequality constraints.
\newblock In \emph{Proceedings of the International Conference on Machine
  Learning}, 937--945. Beijing.

\bibitem[{Garriga and Soroush(2010)}]{garriga2010model}
Garriga, J.L. and Soroush, M. (2010).
\newblock Model predictive control tuning methods: {A} review.
\newblock \emph{Industrial \& Engineering Chemistry Research}, 49(8),
  3505--3515.

\bibitem[{Gelbart et~al.(2014)Gelbart, Snoek, and Adams}]{gelbart2014}
Gelbart, M.A., Snoek, J., and Adams, R.P. (2014).
\newblock {B}ayesian optimization with unknown constraints.
\newblock In \emph{Proceedings of the Thirtieth Conference on Uncertainty in
  Artificial Intelligence}, 250--259.

\bibitem[{Gevers(2005)}]{gevers2005identification}
Gevers, M. (2005).
\newblock Identification for control: From the early achievements to the
  revival of experiment design.
\newblock \emph{European Journal of Control}, 11, 335--352.

\bibitem[{Goldberg et~al.(1998)Goldberg, Williams, and
  Bishop}]{goldberg1998regression}
Goldberg, P.W., Williams, C.K., and Bishop, C.M. (1998).
\newblock Regression with input-dependent noise: A {Gaussian} process
  treatment.
\newblock In \emph{Advances in neural information processing systems},
  493--499.

\bibitem[{Hern\'{a}ndez-Lobato et~al.(2016)Hern\'{a}ndez-Lobato, Gelbart,
  Adams, Hoffman, and Ghahramani}]{JMLR:v17:15-616}
Hern\'{a}ndez-Lobato, J.M., Gelbart, M.A., Adams, R.P., Hoffman, M.W., and
  Ghahramani, Z. (2016).
\newblock A general framework for constrained {B}ayesian optimization using
  information-based search.
\newblock \emph{Journal of Machine Learning Research}, 17(160), 1--53.

\bibitem[{Hoshiya and Saito(1984)}]{hoshiya1984structural}
Hoshiya, M. and Saito, E. (1984).
\newblock Structural identification by extended {Kalman} filter.
\newblock \emph{Journal of Engineering Mechanics}, 110, 1757--1770.

\bibitem[{Khosravi et~al.(2020)Khosravi, Behrunani, Myszkorowski, Smith,
  Rupenyan, and Lygeros}]{khosravi2020performance}
Khosravi, M., Behrunani, V., Myszkorowski, P., Smith, R.S., Rupenyan, A., and
  Lygeros, J. (2020).
\newblock Performance-driven cascade controller tuning with {B}ayesian
  optimization.
\newblock \emph{arXiv preprint arXiv:2007.12536}.

\bibitem[{Kleywegt et~al.(2002)Kleywegt, Shapiro, and Homem-de
  Mello}]{kleywegt2002sample}
Kleywegt, A.J., Shapiro, A., and Homem-de Mello, T. (2002).
\newblock The sample average approximation method for stochastic discrete
  optimization.
\newblock \emph{SIAM Journal on Optimization}, 12, 479--502.

\bibitem[{Lu et~al.(2020)Lu, Kumar, and Zavala}]{lu2020mpc}
Lu, Q., Kumar, R., and Zavala, V.M. (2020).
\newblock {MPC} controller tuning using {B}ayesian optimization techniques.
\newblock \emph{arXiv preprint arXiv:2009.14175}.

\bibitem[{Neumann-Brosig et~al.(2019)Neumann-Brosig, Marco, Schwarzmann, and
  Trimpe}]{neumann2019data}
Neumann-Brosig, M., Marco, A., Schwarzmann, D., and Trimpe, S. (2019).
\newblock Data-efficient autotuning with {B}ayesian optimization: {A}n
  industrial control study.
\newblock \emph{IEEE Transactions on Control Systems Technology}, 28(3),
  730--740.

\bibitem[{Paulson and Mesbah(2018)}]{paulson2018nonlinear}
Paulson, J.A. and Mesbah, A. (2018).
\newblock Nonlinear model predictive control with explicit backoffs for
  stochastic systems under arbitrary uncertainty.
\newblock \emph{IFAC-PapersOnLine}, 51(20), 523--534.

\bibitem[{Piga et~al.(2019)Piga, Forgione, Formentin, and
  Bemporad}]{piga2019performance}
Piga, D., Forgione, M., Formentin, S., and Bemporad, A. (2019).
\newblock Performance-oriented model learning for data-driven {MPC} design.
\newblock \emph{IEEE Control Systems Letters}, 3(3), 577--582.

\bibitem[{Rasmussen(2003)}]{rasmussen2003gaussian}
Rasmussen, C.E. (2003).
\newblock Gaussian processes in machine learning.
\newblock In \emph{Summer School on Machine Learning}, 63--71. Springer.

\bibitem[{Rawlings and Mayne(2009)}]{rawlings2009model}
Rawlings, J.B. and Mayne, D.Q. (2009).
\newblock \emph{Model predictive control: Theory and design}.
\newblock Nob Hill Publishing, Madison, Wisconsin.

\bibitem[{Shahriari et~al.(2015)Shahriari, Swersky, Wang, Adams, and
  De~Freitas}]{shahriari2015taking}
Shahriari, B., Swersky, K., Wang, Z., Adams, R.P., and De~Freitas, N. (2015).
\newblock Taking the human out of the loop: {A} review of {B}ayesian
  optimization.
\newblock \emph{Proceedings of the IEEE}, 104(1), 148--175.

\bibitem[{Snoek et~al.(2012)Snoek, Larochelle, and Adams}]{snoek2012practical}
Snoek, J., Larochelle, H., and Adams, R.P. (2012).
\newblock Practical {B}ayesian optimization of machine learning algorithms.
\newblock In \emph{Advances in Neural Information Processing Systems},
  2951--2959.

\bibitem[{Wang and de~Freitas(2014)}]{wang2014theoretical}
Wang, Z. and de~Freitas, N. (2014).
\newblock Theoretical analysis of {Bayesian optimisation} with unknown
  {Gaussian} process hyperparameters.
\newblock \emph{arXiv preprint arXiv:1406.7758}.

\end{thebibliography}

\end{document}